\documentclass[sigconf]{acmart}

\usepackage{url,hyperref,lineno}
\usepackage{graphicx}
\usepackage{indentfirst}
\usepackage{array,multirow}
\usepackage{times} 
\usepackage{fancyhdr}
\usepackage{booktabs}
\usepackage{multirow}
\usepackage{comment}
\usepackage{siunitx}
\usepackage{multicol}
\usepackage{subcaption}
\usepackage{stfloats}

\newcommand{\ie}{i.\,e.,\ }

\AtBeginDocument{%
 }

\setcopyright{acmlicensed}
\copyrightyear{2026}
\acmYear{2026}
\acmDOI{XXXXXXX.XXXXXXX}
\acmConference[Conference acronym 'XX]{Make sure to enter the correct
 conference title from your rights confirmation email}{June 03--05,
 2018}{Woodstock, NY}
\acmISBN{978-1-4503-XXXX-X/2018/06}

\begin{document}

\title{Hanger Reflex Based Driving Assistance for Drivers with Peripheral Visual Field Defects}

\author{Hailong Liu}
\authornote{Corresponding author.}
\email{liu.hailong@is.naist.jp}
\orcid{0000-0003-2195-3380}
\affiliation{%
 \institution{Nara Institute of Science and Technology (NAIST)}
 \city{Ikoma}
 \state{Nara}
 \country{Japan}
}

\author{Junya Wada}
\email{wada.junya.wh4@naist.ac.jp}
\affiliation{%
 \institution{Nara Institute of Science and Technology (NAIST)}
 \city{Ikoma}
 \state{Nara}
 \country{Japan}
}

\author{Toshihiro Hiraoka}
\email{toshihiro.hiraoka.1970@ieee.org}
\orcid{0000-0003-2195-3380}
\affiliation{%
 \institution{Japan Automobile Research Institute (JARI)}
 \city{Tokyo}
 \country{Japan}
}

\author{Junpei Kuwana}
\email{kuwana@css.risk.tsukuba.ac.jp}
\affiliation{%
 \institution{University of Tsukuba}
 \city{Tsukuba}
  \state{Ibaraki}
 \country{Japan}
}
 
\author{Makoto Itoh}
\email{itoh.makoto.ge@u.tsukuba.ac.jp}
\orcid{0000-0002-4904-1466}
\affiliation{%
 \institution{University of Tsukuba}
 \city{Tsukuba}
  \state{Ibaraki}
 \country{Japan}
}

\author{Takahiro Wada}
\email{t.wada@is.naist.jp}
\orcid{0000-0003-2195-3380}
\affiliation{%
 \institution{Nara Institute of Science and Technology (NAIST)}
 \city{Ikoma}
 \state{Nara}
 \country{Japan}
}

\renewcommand{\shortauthors}{Liu et al.}

\begin{abstract}

Drivers with peripheral visual field defects may fail to notice pedestrians in their peripheral visual field, leading to delayed hazard awareness and increased collision risk. 
This study explores hanger reflex cue (HRC) as a driving assistance method for drivers with peripheral visual field defects, in which mechanical pressure is applied to specific regions of the head to facilitate anticipatory orientation toward potentially risky pedestrians and support safer driving.
In a driving simulator experiment with 15 participants, we compared driving behavior with and without HRC during pedestrian encounters under simulated peripheral visual field defect. 
The results showed that HRC significantly shifted drivers' modal head rotation angle toward the risky pedestrian and significantly increased gaze duration toward that pedestrian. 
Collision occurrence was lower in the w/ HRC condition than in the w/o HRC condition, although the direct effect of HRC on collision occurrence showed only a marginal trend. 
A piecewise structural equation modeling analysis further suggested that HRC may contribute to collision reduction through a sequential pathway from head rotation to gaze allocation and then to collision occurrence. 
These findings provide preliminary evidence that HRC can support anticipatory attention allocation toward peripheral hazards and may offer a promising driving assistance method for drivers with visual field impairment.

\end{abstract}



\maketitle

\section{Introduction}

Progressive visual field defects can occur in several eye diseases, including glaucoma and retinitis pigmentosa. 
These progressive visual field defects may make it difficult for drivers to monitor surrounding traffic environments, thereby compromising driving safety~\citep{lockhart2009driving,vater2022peripheral}.

Glaucoma is an eye disease that can damage the optic nerve and lead to vision loss and blindness, particularly among older adults~\citep{Ramrattan2001}.
It was estimated that approximately 76 million people worldwide would be affected by glaucoma in 2020~\citep{allison2020epidemiology}.
The Tajimi Study~\citep{IWASE20041641} reported that approximately 5\% of 3,870 randomly sampled Japanese individuals aged over 40 years were estimated to have glaucoma.
In many cases, glaucomatous optic nerve damage is associated with increased intraocular pressure~\citep{Ramrattan2001,allison2020epidemiology,IWASE20041641,Weinreb2014}.
Visual field impairment in glaucoma generally begins in the peripheral visual field and gradually progresses toward the central visual field~\citep{Weinreb2014}.

Retinitis pigmentosa is another major eye disease characterized by progressive visual field loss~\citep{xu2020visual}.
Unlike glaucoma, retinitis pigmentosa is primarily an inherited retinal degenerative disease, but it also commonly causes loss of peripheral vision and narrowing of the visual field.
Patients with retinitis pigmentosa often experience night blindness and peripheral visual field loss in the early stage, and their visual field may gradually narrow into so-called tunnel vision as the disease progresses.

Therefore, although glaucoma and retinitis pigmentosa differ in their pathological mechanisms, both diseases can cause progressive peripheral visual field defects while central vision may remain relatively preserved during certain stages of disease progression.
Because central vision is closely related to fine visual acuity and detailed visual perception, patients may not notice these visual changes until the disease has progressed substantially~\citep{IWASE20041641}.
For this reason, patients who have not yet developed central visual field defects may still be able to maintain daily activities, including driving, because the central visual field is primarily responsible for fine visual acuity and detailed visual perception.
However, even when central vision is relatively preserved, peripheral visual field impairment may reduce the ability to monitor surrounding traffic environments~\citep{lockhart2009driving,vater2022peripheral}, which is essential for safe driving.

\subsection{Driving Issues for Drivers with Progressive Peripheral Visual Field Defects}

During driving, timely detection of dynamic changes in the surrounding environment is essential for safe vehicle control.
The peripheral visual field plays an important role in detecting movement and environmental changes outside the current line of sight~\citep{lockhart2009driving,vater2022peripheral}.
This function is particularly relevant to driving because peripheral motion sensitivity may support the detection of potential hazards, such as pedestrians, cyclists, or vehicles approaching from the side~\citep{mckee1984detection,wood2026visual}.
Therefore, progressive peripheral visual field defects may reduce drivers' ability to notice such hazards in a timely manner.
Consistent with this concern, several studies have reported that drivers with visual field defects are at increased risk of vehicle collisions or poorer driving performance compared with drivers with normal vision~\citep{Haymes2008,Tanabe2011,szlyk1992assessment}.
In particular, severe glaucoma patients with visual field impairments in the lower and left portions of the visual field have been reported to have a higher incidence of vehicle collisions in right-driving countries~\citep{Huisingh2015,Kwon2015}.
In addition, drivers with retinitis pigmentosa and varying degrees of peripheral visual field loss showed a higher proportion of accidents than drivers with normal visual field~\citep{szlyk1992assessment}.
Moreover, \citet{Johnson1983} showed that the rate of vehicle collisions among drivers with visual field impairments in both eyes, such as severe glaucoma, was approximately twice that of drivers with normal visual fields.

These findings suggest that peripheral visual field impairments can limit drivers' ability to perceive surrounding traffic environments and respond to potential hazards.
For drivers whose central vision is relatively preserved but whose peripheral visual field is impaired, a pedestrian may be visible when located far ahead in the central visual field, but may become difficult to notice as the vehicle approaches and the pedestrian shifts into the impaired peripheral visual field.
Therefore, even if the driver initially sees the pedestrian at a distance, the driver may fail to perceive the pedestrian's subsequent crossing behavior when it occurs near the vehicle.
This limitation may become particularly critical in time-sensitive situations, such as when a pedestrian or cyclist suddenly appears from the roadside or starts crossing the road near the vehicle.
In such situations, drivers must quickly detect the hazard, orient their head and gaze toward the hazard direction, and initiate an appropriate avoidance response.
Therefore, driving assistance for drivers with peripheral visual field defects should support not only general awareness of the surrounding environment, but also rapid attention allocation and head orientation toward potential hazards.

\subsection{Driving assistance for drivers with visual field defects}

Although relatively few studies have directly examined driving assistance systems specifically for drivers with peripheral visual field defects, several studies have investigated assistance methods for drivers with other types of visual impairment, such as homonymous visual field defects (HVFDs), homonymous field loss, and central vision loss.
These studies are relevant because they also address the problem of delayed hazard detection caused by limited visual information.

First, although many driver assistance systems for drivers with normal vision are based on visual interfaces, such visual cue based approaches may not be sufficient for drivers with visual field defects.
In urgent hazard situations, visual warnings may therefore be missed or may impose additional visual search demands~\citep{xu2024hazard}.
Meanwhile, several studies have explored driver assistance or compensation systems for drivers with visual field defects. 
Visual compensation approaches have been investigated to expand or redirect visual information from the blind field. 
For example, peripheral prism glasses have been evaluated for drivers with hemianopia as a way to improve blind side hazard detection~\citep{houston2018driving,bowers2012pilot}.

Second, auditory assistance has been examined as a means of supporting compensatory scanning behavior. 
\citet{xu2022auditory} investigated auditory reminder cues for drivers with homonymous hemianopia and suggested that such cues may promote proactive blind-side scanning when approaching intersections, which may contribute to faster responses to hazards.

Third, several studies have explored direction-specific vibro-tactile warnings as a non-visual ADAS approach for drivers with visual field impairments. 
\citet{holzl2021driving} developed a seat based vibro-tactile ADAS for drivers with HVFDs, in which brief left or right side vibrations indicated the side from which a pedestrian hazard might appear. 
Similarly, \citet{xu2025directional} proposed a directional vibro-tactile warning system for drivers with central vision loss (CVL) or HVFDs, in which left or right side seat vibrations were delivered when the risk of collision with an approaching pedestrian exceeded a predefined threshold. 
Both simulator studies showed that such side-specific tactile warnings improved blind-side hazard detection and safety driving responses, such as reducing unsafe crossing situations, shortening brake response times, and reducing collision rates. 
These findings suggest that direction-specific tactile warnings can compensate for limited visual hazard detection without adding further visual demand.

However, the above driving assistance systems mainly provide warning information and rely on the driver to voluntarily redirect gaze or head orientation toward the potential hazard.
For drivers with peripheral visual field defects, this voluntary response may be delayed when a pedestrian or other road user is located outside the preserved central visual field.
In anticipatory driving assistance scenarios, it may therefore be useful to support timely attention allocation before the situation becomes critical, rather than only warning the driver after an imminent hazard has emerged.
In this context, an assistance method that can non-visually indicate the relevant direction and physically facilitate head orientation may provide a complementary approach to conventional visual or auditory cue based ADAS.

\subsection{Proposed Hanger Reflex Based Driving Assistance for Drivers with peripheral visual field defects}

\begin{figure}[ht]
  \centering
  \includegraphics[width=1\linewidth]{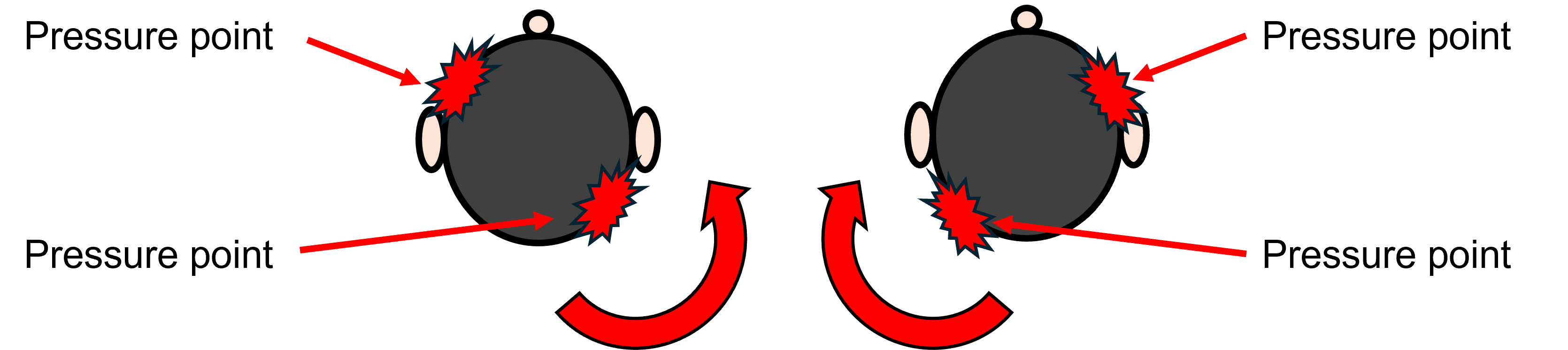}
  \caption{Pressure points for hanger reflex stimulation and the corresponding induced leftward and rightward head rotation directions.}
  \label{fig:hangerpoint}
\end{figure}

To address this issue, this study proposes a hanger reflex based driving assistance method for drivers with peripheral visual field defects.
The hanger reflex is a phenomenon in which mechanical pressure applied to specific regions of the head induces involuntary head rotation~\citep{matsue2008hanger}.
As illustrated in Fig.~\ref{fig:hangerpoint}, applying pressure to the left-front and right-rear regions of the head induces rightward head rotation, while applying pressure to the right-front and left-rear regions induces leftward head rotation.
Previous studies have investigated the hanger reflex as a human-interface technique and developed devices capable of inducing head rotation through mechanical stimulation~\citep{matsue2008hanger,sato2009development,kon2017interpretation}, without relying on visual or auditory information.
Unlike conventional warning systems that only inform drivers of a hazard direction, hanger reflex based stimulation may physically facilitate head orientation toward a target direction.

This property may be particularly useful in risky traffic situations, where drivers with peripheral visual field defects must quickly orient toward potential hazards appearing in the peripheral visual field and initiate an appropriate response.
Thus, hanger reflex based stimulation may provide a novel directional assistance approach that not only indicates the hazard direction but also supports the driver's orienting response.

However, to the best of our knowledge, no previous study has examined whether the hanger reflex can be used as a driving assistance method for drivers with peripheral visual field defects.
It remains unclear whether hanger reflex based stimulation can effectively facilitate head orientation during driving and whether such assistance can improve drivers' responses to potential hazards appearing in the peripheral visual field.

\subsection{Aim of this study}

This study presents an initial feasibility investigation of hanger reflex cue (HRC) as a directional support method for drivers with peripheral visual field defects. 
Specifically, this study has three main aims.
First, we examine whether HRC can still facilitate head rotation toward the risky pedestrian while drivers are actively engaged in vehicle operation.
Second, we examine whether HRC can support drivers with peripheral visual field defects in maintaining traffic safety during pedestrian crossing events by reducing collision occurrence.
We further analyze how HRC affects drivers' head rotation and gaze behavior in this process and exploratorily examine how these behavioral responses are related to collision occurrence.
Finally, we investigate participants' subjective experiences and perceptions of HRC, including its perceived usefulness for safe driving and usability issues such as discomfort and interference with vehicle operation.

\section{Method}

In this experiment, a driving simulator was used to examine how participants responded to hazards under conditions with and without hanger reflex stimulation in situations where a pedestrian crossed or did not cross the road. The effectiveness of the hanger reflex device was evaluated based on drivers’ responses to the hazards.
The experiment received approval from the Research Ethics Committee of 
Nara Institute of Science and Technology (No. 2025-I-49).

\subsection{Participants}

The experiment was conducted with 15 participants (8 males and 7 females with ages of 22--25 years, self-reported) who held a valid Japanese driver's license.
All participants had normal visual fields.
In this experiment, peripheral visual field impairment was simulated by presenting a visual field mask to participants with normal vision.
The entire experiment took approximately 2 hours, and each participant received 2,000 JPY as compensation.

\subsection{Driving Simulator}

\begin{figure}[hb]
  \centering
  \includegraphics[width=1\linewidth, trim=0 0 0 25, clip]{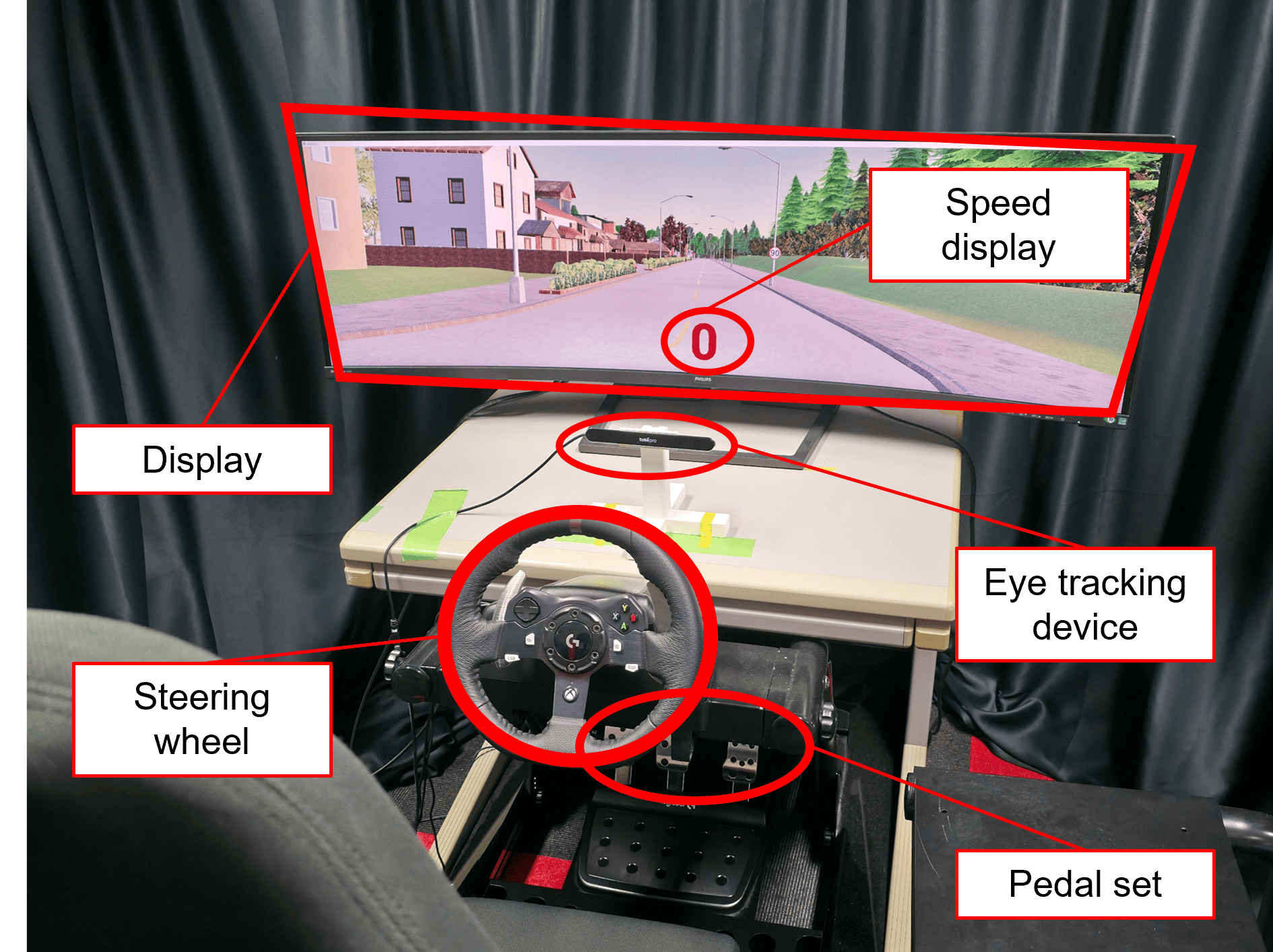}
\caption{Driving simulator (DS) used in the experiment.}
\label{fig:DS}
\end{figure}

The experiment was conducted using a driving simulator system based on CARLA version 0.9.13~\citep{Dosovitskiy17}. 
As shown in Fig.~\ref{fig:DS}, the driving simulator setup consisted of a pedal set and a steering wheel for driving operation input, as well as a 49-inch 32:9 SuperWide curved display with a resolution of 3840 $\times$ 1080 pixels for presenting the simulation graphics (Philips Brilliance 499P).

\subsection{Peripheral visual field defect simulation}

\begin{figure}[!t]
\centering
\includegraphics[width=1\linewidth, trim=0 10 0 8, clip]{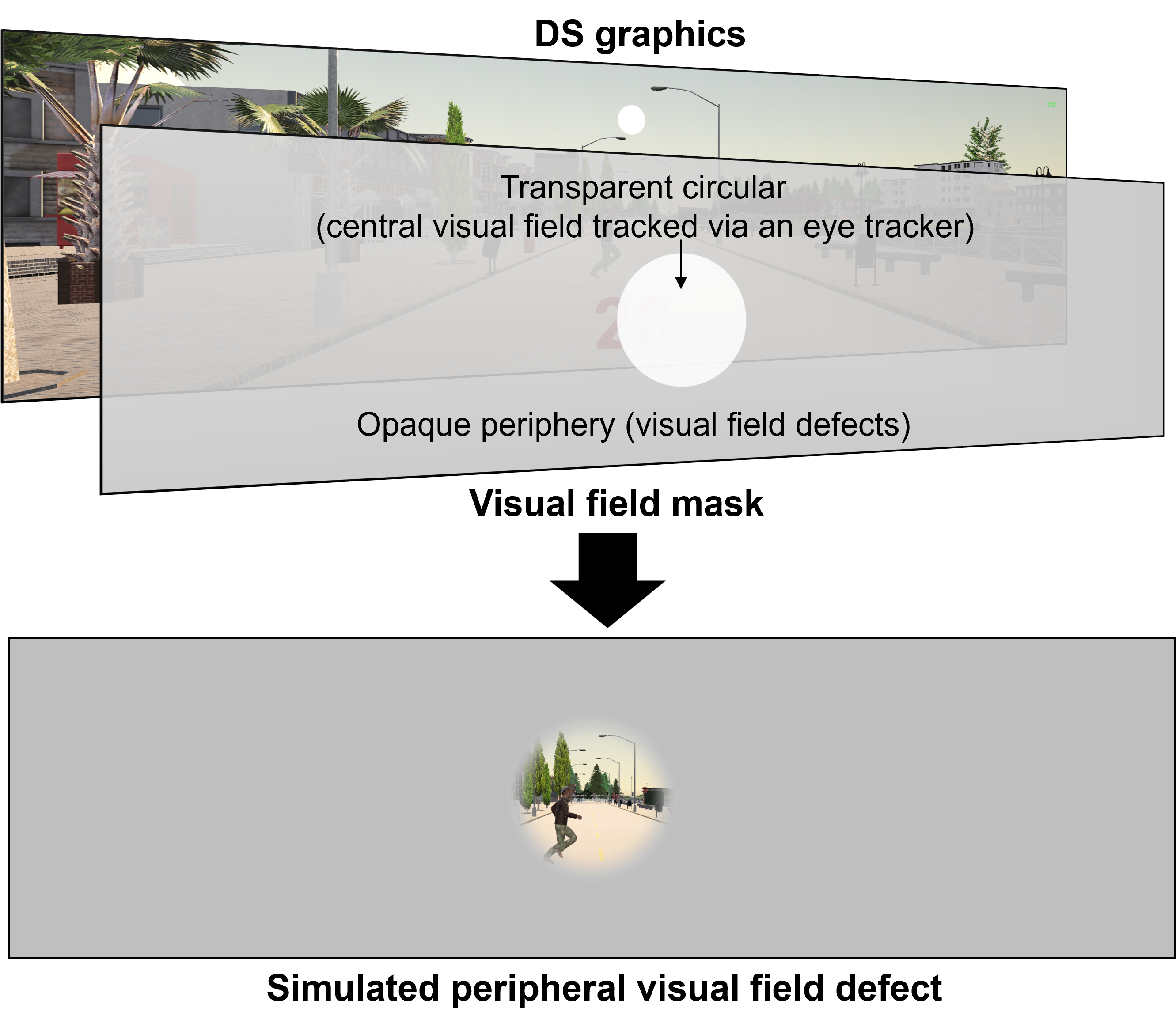}
\caption{Peripheral visual field defect simulation using a visual mask with a central transparent area of approximately $12^\circ$ ($\pm6^\circ$). The transparent area was updated in real time according to the participant's gaze direction.}
  \label{fig:mask}
\end{figure}

With reference to~\citep{liu2023investigating}, a simulation system for peripheral visual field defects was applied in the driving simulator to approximate ``tunnel vision'', as observed in severe bilateral glaucoma and middle to late stage of retinitis pigmentosa.
Specifically, as shown in Fig.~\ref{fig:mask}, a program developed by~\citet{liu2023investigating} was used to project a visual field mask in front of the DS graphics.
A visual mask with a transparent circular area was positioned in front of the DS graphics.
The transparent circular area simulated the preserved central visual field, while the opaque surrounding area simulated peripheral visual field impairment.
The transparent circular area was sized to approximately simulate the central visual field within $12^\circ$ ($\pm6^\circ$).
The position of the center of the transparent circular area was updated in real time according to the participant's gaze point measured by a \textit{Tobii Pro Nano} eye tracker at 60 Hz (see Fig.~\ref{fig:DS}).

\subsection{Hanger reflex device}

\begin{figure}[!t]
  \centering  \includegraphics[width=1\linewidth]{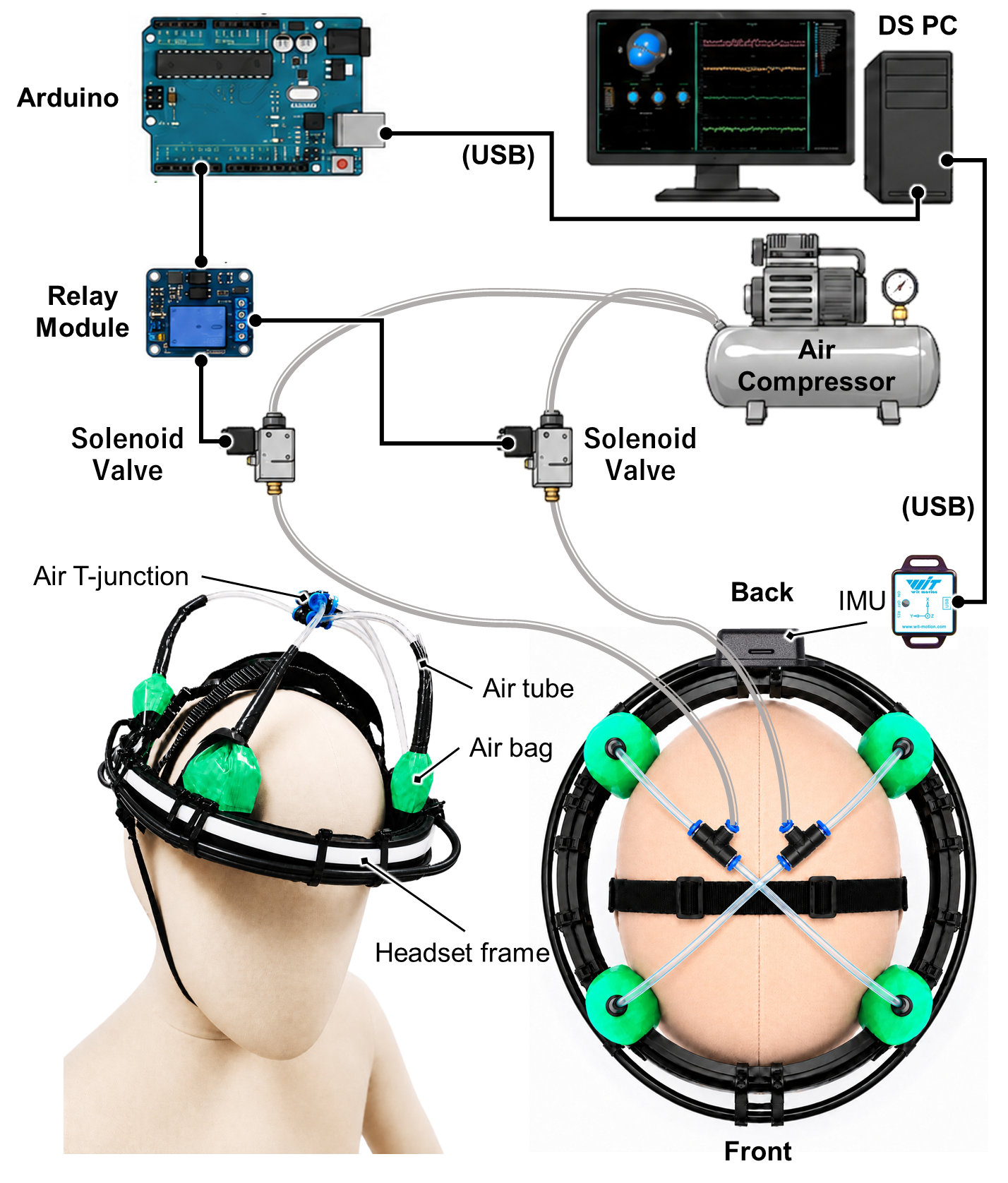}
  \caption{Hanger reflex system used in the experiment.}
  \label{fig:HRD}
\end{figure}

The hanger reflex device was developed with reference to the device design proposed by \citet{kon2018hangerover}.
The device with its control system used in this experiment is shown in Fig.~\ref{fig:HRD}.
It consisted of a tactile stimulation headset worn by the participant, four airbags and an inertial measurement unit (IMU; WitMotion WT901BLE68) mounted on the frame.
The four airbags formed two diagonal pairs around the participant's head, and the two airbags in each pair were connected to the same air T-junction.
Each air T-junction was connected to an electronic air valve.
The two electronic air valves were connected to an air compressor and were controlled by the Arduino through a relay module.
The Arduino was connected to the DS computer via USB and received HRC trigger signals through serial communication.

In this study, hanger reflex stimulation was generated by inflating one diagonal pair of airbags at a time.
Inflating the front-left and rear-right airbags was used to induce rightward head rotation, while inflating the front-right and rear-left airbags was used to induce leftward head rotation (see Fig.~\ref{fig:hangerpoint}).

\subsection{Experimental scenario}

\begin{figure}[t]
  \centering
  \includegraphics[width=1\linewidth]{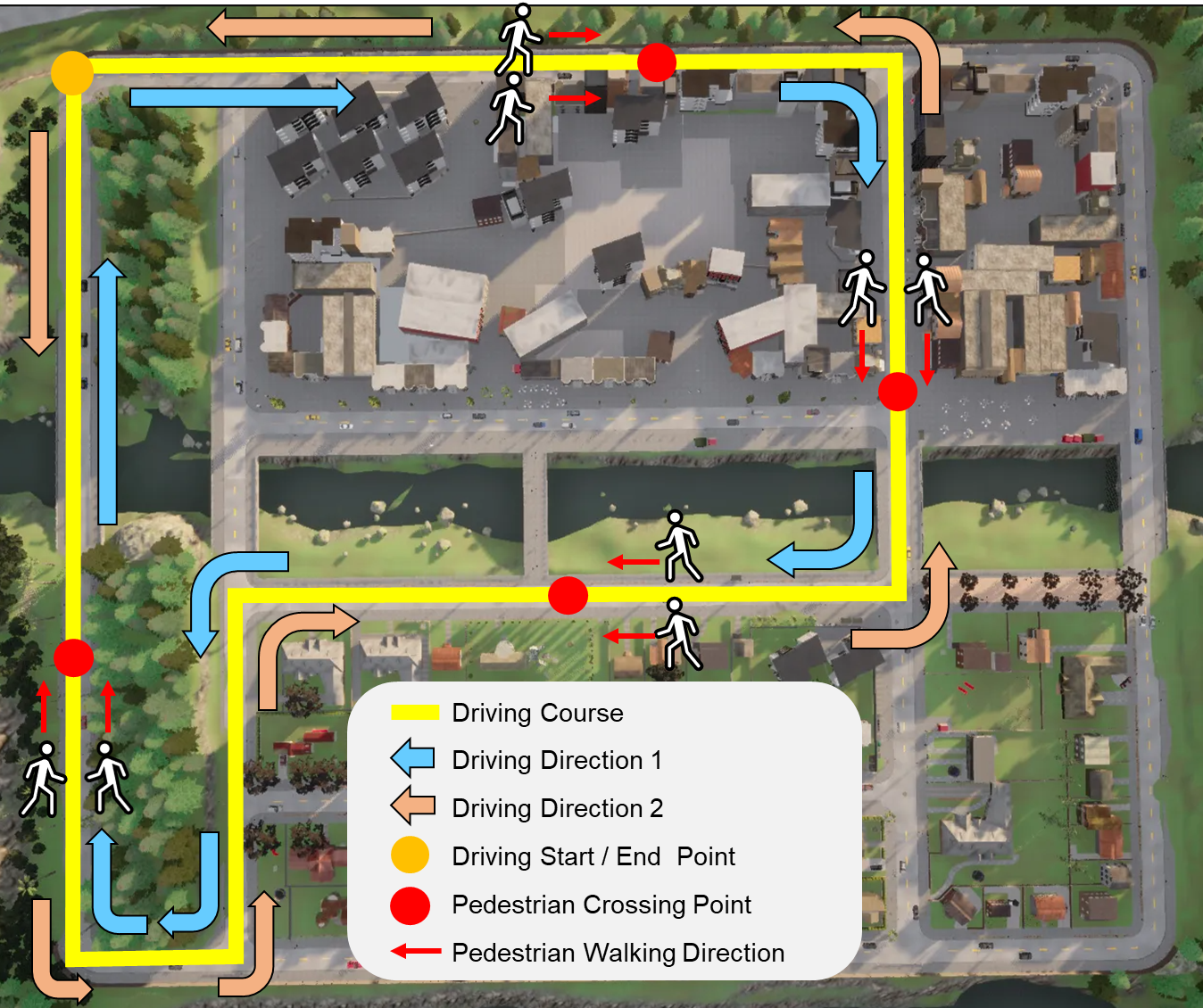}
  \caption{Driving courses and pedestrian encounter scenarios.}
  \label{fig:map}
\end{figure}

The experimental course used in this study is shown in Fig.~\ref{fig:map}. 
The yellow circle indicates the start and end point of the course. 
Each participant drove two laps on the same course in opposite directions, indicated by the blue and light-orange arrows. 
Participants completed one lap under the w/ HRC condition and one lap under the w/o HRC condition, and the order of the two HRC conditions was randomized across participants. 
To help participants follow the predefined course, simulated voice navigation instructions were provided 50~m before each turn.

In each lap, participants encountered pedestrians at four predefined locations along the driving path. 
At each location, two pedestrians were positioned on the left and right sides of the road and walked in the same direction, as indicated by the red arrows in Fig.~\ref{fig:map}. 
Among the four pedestrian encounters, two were randomly assigned as crossing events, in which one pedestrian was randomly selected from either the left or right side to cross the road.
In the remaining two encounters, neither pedestrian crossed the road. 

In the w/ HRC condition, directional stimulation was applied according to the pedestrian situation. 
When one pedestrian crossed the road, stimulation was applied to facilitate the driver's head orientation toward the side of the crossing pedestrian. 
When neither pedestrian crossed, the system selected the pedestrian side with higher potential risk and applied stimulation toward that side. 
The hanger reflex stimulation was therefore designed as an anticipatory directional cue rather than an emergency warning. 
Its purpose was to support timely attention allocation toward a potentially relevant pedestrian before the situation became critical.

In the w/o HRC condition, participants encountered the same types and number of pedestrian situations as in the w/ HRC condition, but drove in the opposite direction. 
To control for the possible influence of wearing the HR device itself, participants also wore the HR device in the w/o HRC condition.
However, the device was not activated, and no HRC stimulation was delivered.

\subsection{Measurement}

\subsubsection{Modal head rotation toward the risky pedestrian}

Head rotation angle was recorded in real time at 100 Hz using an IMU mounted on the HR device. 
In the original IMU coordinate system, positive values indicated rightward head rotation, while negative values indicated leftward head rotation. 
To evaluate whether head rotation was directed toward the risky pedestrian, the sign of the head rotation angle was adjusted according to the side of the risky pedestrian. 
After this transformation, positive values indicated head rotation toward the risky pedestrian, while negative values indicated head rotation away from the risky pedestrian.

The modal head rotation angle was used to evaluate the dominant direction of head rotation during each pedestrian encounter.
Specifically, the head angle at the beginning of each pedestrian encounter was used as the baseline, and the modal head rotation angle within the following 3 s was calculated.
Before calculating the modal value, the continuous head rotation angle was discretized into 1$^\circ$ bins using left-closed and right-open intervals, with each bin represented by its midpoint, e.g., $[-1^\circ, 0^\circ)$ as $-0.5^\circ$ and $[0^\circ, 1^\circ)$ as $0.5^\circ$.

\subsubsection{Gaze duration toward the risky pedestrian}

Participants' gaze behavior was recorded during each pedestrian encounter, from the moment the vehicle was 50 m away from the pedestrians until it passed them.
As shown in Fig.~\ref{fig:mask}, the transparent circular area of the gaze-contingent visual mask was treated as the participant's simulated central visual field.
The participant was considered to be gazing at the risky pedestrian when the pedestrian was included within this area.
Gaze duration toward the risky pedestrian was calculated as the cumulative duration of these gaze episodes during the encounter.

\subsubsection{Collisions in pedestrian-crossing events}

To investigate whether HRC can reduce collision risk between drivers with peripheral visual field defects and pedestrians, collision occurrences were recorded during pedestrian crossing events. 
Each participant encountered pedestrians in eight events in total across the w/o HRC and w/ HRC conditions, four of which involved pedestrian crossing and were included in the collision occurrence analysis.

\subsubsection{Post-experiment subjective evaluation}

The post-experiment subjective evaluation consisted of the following three items:
\begin{itemize}
  \item Q1. Compared with the no-HRC condition, did HRC make you feel uncomfortable during vehicle operation?

  \item Q2. Compared with the no-HRC condition, did HRC interfere with your vehicle operation?

  \item Q3. Compared with the no-HRC condition, did you think HRC was useful for safe driving?
\end{itemize}
Responses were provided using a four-point Likert scale,
\ie ``Agree'', ``Slightly agree'', ``Slightly disagree'', and ``Disagree''.

\subsection{Pathway analysis using piecewise structural equation modeling}

To examine the hypothesized behavioral pathway by which HRC may contribute to collision avoidance, we conducted a piecewise structural equation modeling (pSEM) analysis. 
The hypothesized sequential pathway assumed that HRC would first facilitate head rotation toward the risky pedestrian, that greater head rotation toward the risky pedestrian would increase gaze duration toward the risky pedestrian, and that longer gaze duration would reduce the likelihood of collision occurrence. 
Thus, the following pathway was tested: HRC $\rightarrow$ head rotation toward the risky pedestrian$\rightarrow$ gaze duration toward the risky pedestrian $\rightarrow$ collision occurrence.

The pSEM consisted of three component mixed-effects models. 
First, head rotation angle was modeled as a function of HRC. 
Second, gaze duration toward the risky pedestrian was modeled as a function of head rotation angle. 
Third, collision occurrence was modeled as a function of gaze duration. 
Participant ID was included as a random intercept in all component models to account for repeated observations within participants. 
Linear mixed-effects models were used for head rotation angle and gaze duration, while a generalized linear mixed-effects model with a binomial distribution and logit link function was used for collision occurrence.
The component models were specified as follows:
\begin{equation}
\begin{aligned}
\text{Head rotation} \quad &\sim \quad \text{HRC} + (1|\text{ID}), \\
\text{Gaze duration} \quad &\sim \quad \text{Head rotation} + (1|\text{ID}), \\
\text{Collision} \quad &\sim \quad \text{Gaze duration} + (1|\text{ID}).\nonumber
\end{aligned}
\end{equation}

The pSEM was implemented using the \texttt{piecewiseSEM} package in R. 
Model fit was evaluated using the tests of directed separation, the chi-square statistic, and Fisher's $C$ statistic. 
A non-significant Fisher's $C$ was interpreted as indicating no significant evidence of poor model fit.

\begin{figure*}[!b]
  \centering
  \begin{minipage}{0.32\linewidth}
    \centering
    \includegraphics[width=\linewidth, trim=0 13 0 17, clip]{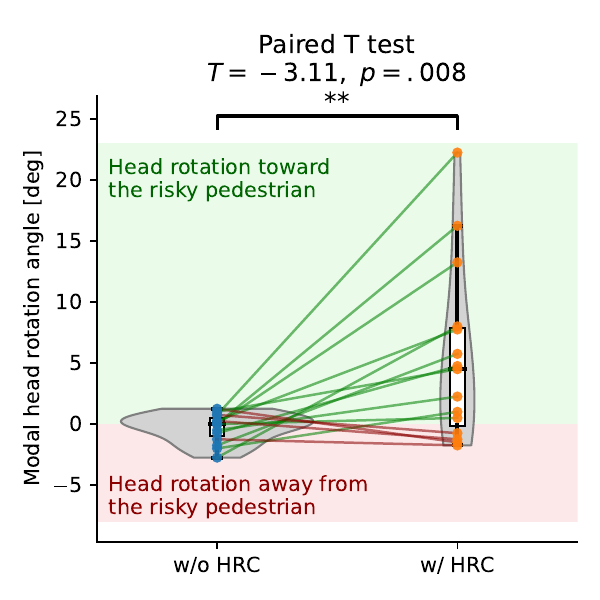}
    \captionof{figure}{Modal head rotation toward the risky pedestrian under the w/o HRC and w/ HRC conditions.}
    \label{fig:head_angle_HRC}
  \end{minipage}
  \hfill
  \begin{minipage}{0.32\linewidth}
    \centering
    \includegraphics[width=\linewidth, trim=0 13 0 17, clip]{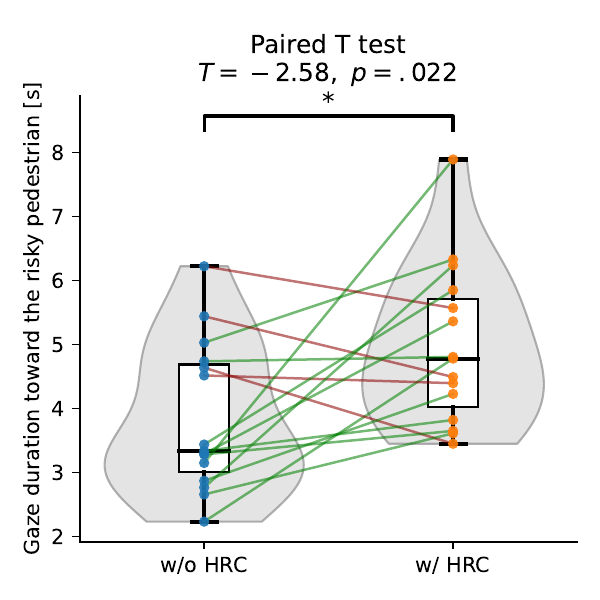}
    \captionof{figure}{Driver's gaze duration toward the risky pedestrian under the w/o HRC and w/ HRC conditions.}
    \label{fig:Gaze}
  \end{minipage}
   \hfill
  \begin{minipage}{0.32\linewidth}
    \centering
    \includegraphics[width=\linewidth, trim=0 0 10 0, clip]{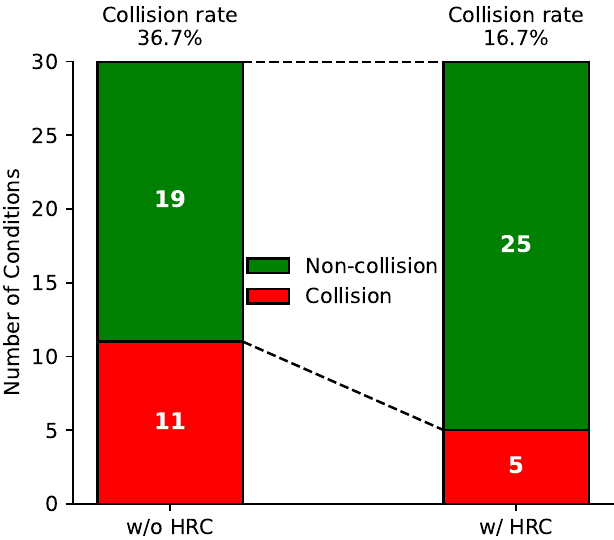}
    \captionof{figure}{Collisions in pedestrian-crossing events under the w/o HRC and w/ HRC conditions.}
    \label{fig:collision}
  \end{minipage}
\end{figure*}

\section{Results}

\subsection{Modal head rotation toward the risky pedestrian}

Fig.~\ref{fig:head_angle_HRC} shows the modal head rotation angle under the w/o HRC and w/ HRC conditions using paired violin plots.
Positive values, shown by the light green background, indicate head rotation toward the risky pedestrian, while negative values, shown by the light red background, indicate rotation away from the risky pedestrian.
Individual paired lines represent paired data from the same participant.
Green lines indicate an increase in modal head rotation angle under the w/ HRC condition compared with the w/o HRC condition, while red lines indicate a decrease.

The normality of the paired differences was examined using the Shapiro--Wilk test.
No significant deviation from normality was observed, $W = 0.923$, $p = .214$.
Therefore, a paired $t$-test was conducted.
The paired $t$-test revealed a significant difference between the w/o HRC and w/ HRC conditions, $t(14) = -3.11$, $p = .008$.
The modal head rotation angle shifted in the positive direction under the w/ HRC condition compared with the w/o HRC condition, indicating that HRC significantly promoted head rotation toward the risky pedestrian.

\subsection{Gaze duration toward the risky pedestrian}

As shown in the paired violin plots in Fig.~\ref{fig:Gaze}, the distribution of the driver's gaze duration toward the risky pedestrian shifted upward in the w/ HRC condition compared with the w/o HRC condition.
The individual paired lines further indicate that most participants increased their gaze duration toward the risky pedestrian when HRC was applied.

The normality of the paired differences was examined using the Shapiro--Wilk test.
No significant deviation from normality was observed, $W = 0.962$, $p = .728$.
Therefore, a paired $t$-test was conducted.
The paired $t$-test revealed a significant difference between the two conditions, $t(14) = -2.58$, $p = .022$, indicating that HRC significantly increased drivers' gaze duration toward the risky pedestrian.

\subsection{Collisions in pedestrian-crossing events}

As shown in Fig.~\ref{fig:collision}, the collision rate was lower in the w/ HRC condition than in the w/o HRC condition. 
Specifically, collisions occurred in 5 out of 30 trials in the w/ HRC condition, corresponding to a collision rate of 16.7\%, while collisions occurred in 11 out of 30 trials in the w/o HRC condition, corresponding to a collision rate of 36.7\%.

A mixed-effects logistic regression analysis was then conducted with collision occurrence as the dependent variable, HRC condition as the fixed effect, and participant ID as a random intercept, \ie $\text{Collision} \sim \text{HRC} + (1|\text{ID})$. 
The model resulted in a singular fit, with the random-intercept variance estimated as zero. Therefore, the random effect did not explain additional variance beyond the fixed effect, and the fixed-effect estimate was interpreted as equivalent to that from a standard logistic regression model.
As shown in Table~\ref{tab:logistic_collision}, the estimated coefficient for the w/ HRC condition was negative ($\beta = -1.063$), with an odds ratio of 0.345. 
This result indicates that collision odds tended to be approximately 65.5\% lower when HRC was provided.
However, this tendency did not reach statistical significance at the .05 level, although a marginal trend was observed ($p = .086$).
This may be partly due to the limited number of participants and collision events.

\begin{table}[!b]
\centering
\caption{Logistic regression results for the effect of HRC on collision occurrence.}
\label{tab:logistic_collision}
\resizebox{\linewidth}{!}{%
\footnotesize
\begin{tabular}{lrrrrr@{\hspace{2pt}}l}
\toprule
Source & $\beta$ & \textit{OR} & \textit{SE} & \textit{z} & \multicolumn{2}{c}{\textit{p}} \\
\midrule
Intercept & -0.547 & 0.579 & 0.379 & -1.443 & .149 & \\
HRC & -1.063 & 0.345 & 0.619 & -1.716 & .086 & $\dagger$ \\
\midrule
\multicolumn{7}{l}{Reference condition: w/o HRC. OR = odds ratio.} \\
\multicolumn{7}{l}{ $\beta$ = regression coefficient estimate.$\dagger p < .10$.} \\

\end{tabular}%
}
\end{table}

\subsection{Post-experiment subjective evaluation}

\begin{figure*}[htb]
  \centering  \includegraphics[width=0.65\linewidth, trim=0 10 0 5,clip]{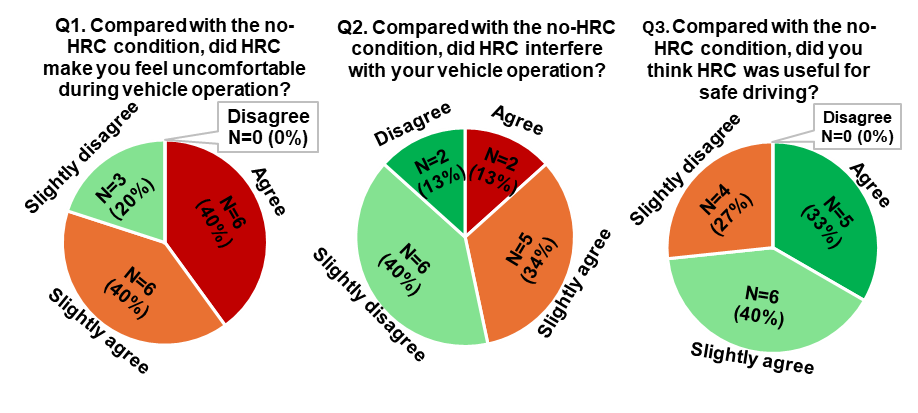}
  \caption{Post-experiment subjective evaluation results}
  \label{fig:Q}
\end{figure*}

\begin{figure*}[h!t]
  \centering
  \includegraphics[width=0.85\linewidth]{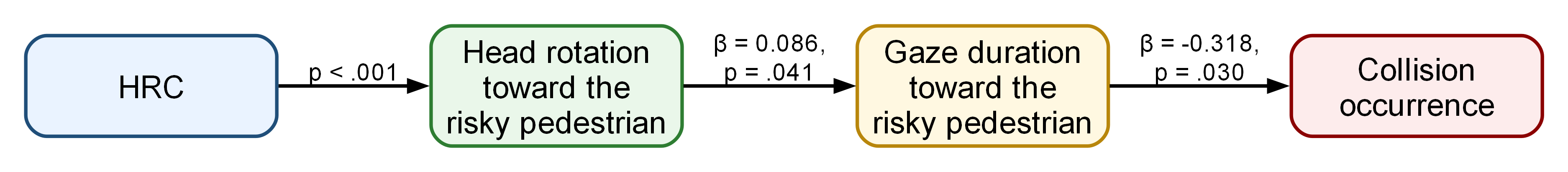}
  \caption{Piecewise structural equation model (pSEM) showing the pathway from HRC to collision occurrence through head rotation and gaze duration toward the risky pedestrian.}
  \label{fig:pSEM}
\end{figure*}

\begin{table*}[h!t]
\centering
\caption{Results of the piecewise structural equation model examining the effects of HRC on collision occurrence through head rotation toward the risky pedestrian and gaze duration.}
\label{tab:psem_results}
\resizebox{\textwidth}{!}{%
\footnotesize
\begin{tabular}{lllrrrrr@{\hspace{2pt}}l}
\toprule
Model type & Model formula & Term & $\beta$ & SE & DF & Crit. value & \multicolumn{2}{c}{$p$} \\
\midrule
LMM 
& $\text{Head rotation} \sim \text{HRC} + (1|\text{ID})$ 
& HRC & -- & -- & 1.00 & 16.318 & $<$.001 & *** \\
& & w/o HRC & -0.767 & 1.310 & 36.07 & -0.585 & .562 & \\
& & w/ HRC & 6.133 & 1.310 & 36.07 & 4.681 & $<$.001 & *** \\
\midrule
LMM
& $\text{Gaze duration} \sim \text{Head rotation} + (1|\text{ID})$ 
& Head rotation & 0.086 & 0.041 & 58.00 & 2.096 & .041 & * \\
\midrule
GLMM
& $\text{Collision} \sim \text{Gaze duration} + (1|\text{ID})$ 
& Gaze duration & -0.318 & 0.147 & 60.00 & -2.171 & .030 & * \\
\bottomrule
\multicolumn{9}{l}{Model fit of the pSEM: $\chi^2 = 6.953$, $p = .073$; Fisher's $C = 9.345$, $p = .155$. * $p < .05$, *** $p < .001$.} \\
\multicolumn{9}{l}{LMM = linear mixed-effects model. GLMM = generalized linear mixed-effects model. $\beta$ = regression coefficient estimate.}\\
\multicolumn{9}{l}{ SE = standard error. Crit. value indicates the $t$ value for LMMs and the Wald $z$ value for GLMM.} \\
\end{tabular}%
}
\end{table*}

As shown in Fig.~\ref{fig:Q}, 80.0\% of participants answered either ``Agree'' or ``Slightly agree'' for Q1, indicating that many participants felt some discomfort when HRC was provided during vehicle operation. 
For Q2, 46.7\% of participants answered either ``Agree'' or ``Slightly agree'', while 53.3\% answered either ``Slightly disagree'' or ``Disagree'', suggesting that participants' perceptions of interference were mixed.
For Q3, 73.3\% of participants answered either ``Agree'' or ``Slightly agree'', indicating that most participants perceived HRC as useful for safe driving.

\subsection{Pathway Analysis of HRC Effects on Collision}

The pSEM results were consistent with the hypothesized sequential pathway from HRC to collision occurrence through head rotation angle and gaze duration, as illustrated in Fig.~\ref{fig:pSEM}. 
As shown in Table~\ref{tab:psem_results}, the pSEM did not provide sufficient evidence to reject the hypothesized structural model, $\chi^2 = 6.953$, $p = .073$, and Fisher's $C = 9.345$, $p = .155$. 
This indicates that the proposed set of relationships showed an acceptable level of consistency with the observed data.
In the pSEM, HRC significantly affected head rotation angle, $p < .001$, and head rotation angle significantly predicted gaze duration toward the risky pedestrian, $\beta = 0.086$, $p = .041$. 
Furthermore, gaze duration significantly predicted collision occurrence, $\beta = -0.318$, $p = .030$, suggesting that longer visual attention toward the risky pedestrian was associated with a lower likelihood of collision.

\section{Discussion}

\subsection{Effectiveness of the hanger reflex in an active driving task}

Although the hanger reflex has been applied in VR~\citep{matsue2008hanger} and walking navigation~\citep{kon2017interpretation}, its effectiveness during driving has not been sufficiently examined. 
Because driving requires continuous attention and vehicle control, it is unclear whether hanger reflex stimulation can still induce head orientation during an active driving task. 
A key finding of this study, based on the comparison between the w/o HRC and w/ HRC conditions, is that HRC promoted drivers' head rotation toward the risky pedestrian even during active driving (see Fig.~\ref{fig:head_angle_HRC}), suggesting that the directional stimulation remained effective in this highly demanding context.

\subsection{Effects of HRC on driving safety under simulated peripheral visual field defect}

In pedestrian-crossing situations under the simulated peripheral visual field defect, the comparison between the w/o HRC and w/ HRC conditions showed a marginal trend toward a lower collision rate when HRC was provided (see Table~\ref{tab:logistic_collision}).
Specifically, HRC significantly promoted head rotation toward the risky pedestrian (see Fig.~\ref{fig:head_angle_HRC}) and also significantly increased gaze duration toward that pedestrian (see Fig.~\ref{fig:Gaze}). 
The pSEM results (see Fig.~\ref{fig:pSEM}) further supported this process, showing that HRC was associated with head rotation, head rotation was associated with gaze duration, and gaze duration was associated with collision occurrence. 
In summary, these results suggest that HRC may contribute to collision reduction by facilitating drivers' head orientation and visual attention toward the crossing pedestrian.

\subsection{Subjective evaluations for the HRC}

Although HRC tended to reduce collision occurrence in the experiment, the post-experiment subjective evaluation revealed both strong perceived usefulness and important usability issues. 
As shown in Fig.~\ref{fig:Q}, for Q3, the majority of participants reported that HRC was useful for improving driving safety in the driving task with the simulated peripheral visual field defect, suggesting that participants generally recognized the potential benefit of the proposed HRC based driving assistance method. 
However, most participants also reported feeling uncomfortable when using the HR device (Q1), and approximately half of the participants reported that the device interfered with their driving operation (Q2). 
These findings indicate that HRC has potential as a safety-support method, but further improvement is needed to enhance comfort and reduce interference with driving operation.

The interview comments suggest that this discomfort and interference were mainly related to the involuntary change in head direction and visual field induced by the device. 
For example, one participant stated, ``My face moved regardless of my own intention, and my gaze was shifted.'' 
Another participant commented, ``Because my visual field was changed while driving, I could not concentrate on driving.'' 
These comments indicate that the semi-forced nature of HRC may create a sense of unnaturalness or reduced voluntary control, even though it can support orientation toward risky pedestrians.
Therefore, future work should further consider the usability and comfort of HR based driving assistance. 
In particular, drivers may need a longer familiarization period to adapt to the use of the HR device during driving. 
In addition, the stimulation intensity, timing, and presentation method should be optimized so that HRC can provide sufficient directional support while minimizing discomfort and interference with vehicle operation.

\subsection{Limitations and Future Work}

First, the experiment was conducted with only 15 participants.
Although the results provide preliminary evidence for the potential effectiveness of hanger reflex cues, the limited sample size restricts the generalizability of the findings. 
Future studies should recruit a larger and more diverse participant sample to further validate the robustness of the results.

Second, peripheral visual field defects were simulated using a gaze-tracking based visual field mask, and all participants were young people with normal vision. Therefore, their visual, cognitive, and driving abilities may differ from those of actual drivers with peripheral visual field defects.
In addition, only one type of visual field mask was used in this experiment, while the extent and pattern of visual field defects can vary among individuals with glaucoma and retinitis pigmentosa. 
Therefore, the results may differ from those of actual drivers with peripheral visual field defects.
Future work should involve participants with glaucoma or retinitis pigmentosa to examine the practical applicability of the proposed method.

Third, this study did not directly compare HRC with other assistance methods, such as auditory warnings~\citep{xu2022auditory} or vibration based haptic cues~\citep{holzl2021driving,xu2025directional}.
Future studies should compare these cue modalities under the same experimental conditions to clarify the relative advantages and limitations of HRC assistance.

\section{Conclusion}
This study examined the feasibility of using hanger reflex based directional cueing as a driving assistance method for drivers with peripheral visual field defects. 
The results of the simulated experiment provide preliminary evidence that HRC can facilitate drivers' orienting behavior toward potential pedestrian-crossing risks. 
Specifically, HRC significantly shifted the modal head rotation angle toward the risky pedestrian and increased gaze duration toward that pedestrian. 
In addition, the collision rate in pedestrian-crossing events was lower under the w/ HRC condition than under the w/o HRC condition, although the direct effect of HRC on collision occurrence showed only a marginal trend. 
These findings suggest that HRC may support anticipatory attention allocation in pedestrian-crossing situations.


\begin{acks}
This work was supported by the Japan Society for the Promotion of Science (JSPS) KAKENHI Grant Number 24H00361, Japan.
The authors used OpenAI ChatGPT (GPT-5.3) for English proofreading and take full responsibility for the final content.
\end{acks}

\section*{CRediT author statement}

\textbf{Hailong~Liu}: Conceptualization, Methodology, Software, Validation, Formal Analysis, Visualization, Project administration, Funding acquisition,  Writing - Original Draft \& review \& editing.

\textbf{Junya~Wada}: Methodology, Software, Investigation, Data Curation, Formal Analysis, Writing - review \& editing.

\textbf{Toshihiro~Hiraoka}: Methodology, Funding acquisition, Writing - review \& editing.

\textbf{Junpei~Kuwana}: Software, Writing - review \& editing.

\textbf{Makoto~Itoh}: Software, Funding acquisition, Writing - review \& editing.

\textbf{Takahiro~Wada}: Methodology, Funding acquisition, Writing - review \& editing.

\bibliographystyle{ACM-Reference-Format}
\bibliography{sample-base}

\end{document}